# Discovery of a Paired Gaussian, Long-Tailed Distribution of Potential Energies in Nano Glasses

D. M. Zhang[a], D. Y. Sun[a,b] and X. G. Gong[b,c]

[a]Engineering Research Center for Nanophotonics & Advanced Instrument (MOE), School of Physics and Electronic Science, East China Normal University, Shanghai 200241, China
[b]Key Laboratory for Computational Physical Sciences (MOE), State Key Laboratory of Surface Physics, Department of Physics, Fudan University, Shanghai 200433, China
[c]Collaborative Innovation Center of Advanced Microstructures, Nanjing 210093, China

## Abstract

It is generally believed that the intrinsic properties of glasses are intimately related to potential energy landscapes (PELs). However, little is known about the PELs of glasses below the glass transition temperature ($T_g$). Taking advantage of lower potential energy barriers in nano systems, we have systematically investigated the dynamics behavior of two nano glasses, $Al_{43}$ and $Al_{46}$. Structure transformation is identified in our pure molecular-dynamics simulation far below $T_g$, which manifests the existence of metabasins in PELs, at least for nano glasses. Surprisingly, we find that the distribution of potential energies shows a paired-Gaussian and long-tailed distribution at temperatures below and approaching $T_g$, correspondingly the distribution of the α-relaxation time exhibits an exponential-like decay. In contrast to the Gaussian distribution of energy in typical liquids and solids, the unexpected distribution may reflect the intrinsic feature of nano glasses. Associated with the exponential-like distribution of the α-relaxation time, the stretched-exponential structural relaxation is found, and the maximum stretched behavior appears around $T_g$. Despite our studies focused on nano glasses, the current finding may shed light on future studies of bulk glasses.

# Introduction

The perspective of the potential energy landscape (PEL) to explore the nature of glasses came from Goldstein's seminal paper,[1] in which a direct connection between glass transitions and PELs was constructed. Since then, this picture has been developed extensively for supercooled liquids and glasses, and an overwhelming wealth of literature on PELs has been published.[2-6] At present, it is generally recognized that the existence of metabasins (MBs) is a significant feature of PELs for glasses.[5-8] A basin is a region of minima in PELs with similar potential energies, and each minimum corresponds to an inherent structure (IS), *i.e.*, the locally stable configuration.[9] Basins are thought to be organized into groups, which form MBs.[8]

MBs in PELs provide a reasonable scenario for understanding supercooled liquids and glasses.[3] It is believed that the intra-MB transition may involve the rearrangement of a relatively small number of particles,[10,11] which gives rise to the *β*-relaxation (or so-called fast *β*-process). Inter-MB transitions are believed to involve complete rearrangement of local structure, which is closely related to the so-called *α*-relaxation.[6,12-14] The intra- or inter-MB transitions, which are the most fundamental relaxation processes in glass, may also correlate to spatial dynamic heterogeneity.[10,15-20]

PELs supply a physical basis for establishing various thermodynamic models of glassy states. As early as half a century ago, Anderson, Halperin, and Varma,[21] and Phillips[22] had proposed the two-level model to explain the low-temperature anomalous specific heat of glasses. In recent years, a few alternative simplified models have been proposed for various purposes, such as, but not limited to, the trap model,[23] the random energy model,[24] the Gaussian model,[25] the logarithmic model,[26] the double Gaussian model,[27] the mosaic model,[28,29] the constrained excitations models[30-32] and so on.[33-39] All of these simplified models share a common point that PELs are assumed to have a specific *structure*, in which the Gaussian and random distribution of potential energies are widely adopted.

Undoubtedly, the reliability of such simplified thermodynamic models depends on whether the intrinsic feature of PELs is properly abstracted. Although many studies focused on PELs,[40-52] up to date, few studies directly '**saw**' a true PEL of glasses below the glass transition temperature ($T_g$). In most works, PELs of glasses below $T_g$ were mapped from a series of local minima approached by freezing the atomic motion in a higher temperature liquid. At a high enough temperature, the potential energy distribution will eventually be Gaussian. By kinetic "quenching", the Gaussian distribution should remain. However, it is not clear how the real PEL of a kinetic-frozen glass relates to that of a thermally cooled one.

For bulk glasses the structural relaxation process is extremely slow as $T_g$ is approached, which has exceeded the affordable timescale of conventional dynamic simulations.[53] Moreover, direct experimental evidence is even more scarce.[54] This leads to the current situation, although the basic scenario of PELs is ready,[3] the general features of PELs below $T_g$ is insufficient.

In recent decades, a new trend on glass research, namely low-dimensional glasses, emerges, in which the representative systems are thin films[55] and nano glasses.[56,57] Although these low-dimensional glasses may have its own unique properties, they all share the common characteristics of glassy state, such as exhibiting the glass transition and relaxation behavior. Previous studies have shown that, even for a single disordered nanoclusters, it can undergo a very broad transition in their dynamics and thermodynamics that resembles glass-formation in bulk materials in many ways.[58-60] More importantly, the structural relaxation time of nano systems containing dozens of atoms could be significantly shorter.[59-65] Many unreachable relaxation processes in bulk glass can be observed in nanoscale counterparts in conventional computer simulations. Taking this advantage into account, nano glasses could be an excellent system to explore PELs below $T_g$.

In this paper, by performing molecular dynamics (MD) simulations in the microsecond scale, we have directly observed the transition among MBs in PELs. More importantly, we find that the potential energy shows a paired-Gaussian and long-tailed distribution, associated with an exponential distribution for the $\alpha$-relaxation time.

## Computational Details

The interaction between Al atoms is described by the glue potential[66]. Previous studies showed that both $Al_{43}$ and $Al_{46}$ have disordered structures in their ground states[58,59,67,68], and melt or solidify with a typical glass-like transition with $T_g = 510$ K and 527 K respectively. In this work, by using MD simulations, we have explored the PEL of the two nano glasses, $Al_{43}$ and $Al_{46}$.

Like the Kob-Andersen model,[69] the current systems can undergo a glass transition with an extremely low cooling rate, and can be considered to be in an equilibrium state at any temperature. At each temperature of interest, a 10 ns simulation for initial relaxation is performed, followed by a 6 μs simulation. To calculate the relaxation time and characterize the structural change, the self-intermediate scattering function (SISF) and mean square displacement (MSD) are calculated. SISF is defined as[69]

$$F_s(q = q_{\max}, t) = N^{-1} \sum_{i=1}^{N} \exp\{i\boldsymbol{q} \cdot [\boldsymbol{r}_i(0) - \boldsymbol{r}_i(t)]\}, \qquad (1)$$

where $\boldsymbol{r}_i(0)$ and $\boldsymbol{r}_i(t)$ are the coordinates of the $i$-th atom at time 0 and $t$, respectively; and $q_{\max}$ is the wave vector at which the static structure factor $S(q)$ reaches the first main peak. In the present work, the value of $q_{\max}$ barely changes within the temperature range of interest, and is set as 2.84 Å$^{-1}$. After simulation, the 6-μs data is divided into 1500 intervals, each with a duration of 4 ns. With SISF being calculated for each 4-ns interval, one could statistically examine the distribution and fluctuation of the structural relaxation time.

To show correspondence between cooperative rearrangement and the evolution of energy in time, the overlap function $Q(t)$ is calculated:[70-72]

$$Q(t) = \frac{1}{N} \sum_{i=1}^{N} \sum_{j=1}^{N} \theta(a - |\boldsymbol{r}_i(0) - \boldsymbol{r}_j(t)|), \qquad (2)$$

where $\theta(x)$ is the Heaviside function, *i.e.*, $\theta(x) = 1$ if $x \geq 0$, and $\theta(x) = 0$ if $x < 0$. In this work, we set $a = 0.3a_0$ where $a_0$ is the lattice constant. In order to eliminate the effect of exchange between atom positions, all atoms are cycled in the second summation. The overlap function $Q(t)$ reflects the time evolution of structural similarity of the system, its jump is the result of the cooperative rearrangement of atoms, and the larger the change is, the more atoms are involved.

In order to eliminate thermal fluctuations in potential energies, the short-time average of potential energy ($E_{\text{ave}}$) is calculated as[59]

$$E_{\text{ave}}(t; \Delta t) = \frac{1}{\Delta t} \int_0^{\Delta t} E_{\text{pot}}(t + \tau) \, d\tau, \tag{3}$$

where $E_{\text{pot}}(t)$ is the potential energy at time $t$, and $\Delta t$ is a short time interval. When $\Delta t$ is comparable to the correlation time, the fluctuation caused by atomic vibrations will be first smoothed out. If $\Delta t$ increases further, namely comparable to the $\beta$-relaxation time, the change in $E_{\text{ave}}(t; \Delta t)$ could be a reflection of transitions among MBs. Roughly speaking, $E_{\text{ave}}(t; \Delta t)$ should mainly contain the structural relaxation information with the relaxation time longer than $\Delta t$. If $E_{\text{ave}}(t; \Delta t)$ is synchronized with a notable change in $Q(t)$, we can conclude that the cooperative rearrangement corresponds to a jump from one PEL minimum to another.

It is well-known that nano systems may inevitably have so-called surface effects. However previous study[59] has shown that, the inner and surface atoms have comparable diffusive behavior during the glass transition even below $T_g$. In other words, the glass transition in nano glasses cannot be simply considered as surface effects. Considering that nano and bulk glasses share many same or similar thermodynamic properties[68], the current research will shed light on future studies about bulk or film glasses.

## Result and Discussion

### 1) Non-Gaussian Distribution of Potential Energies

With the decrease of temperature, the distribution of potential energy of both nano glasses presents a more and more obvious non-Gaussian feature. Fig. 1(a) and (b) depict the distribution of potential energy for $Al_{43}$ and $Al_{46}$, respectively, where the dashed line fits to a Gaussian distribution. To quantitatively describe the deviation, the skewness ($S$) and excess kurtosis ($K$) are calculated (inset of Fig. 1). $S$ and $K$ are defined as, $S =$

$\langle(\frac{x-\mu}{\sigma})^3\rangle$ and $K = \langle(\frac{x-\mu}{\sigma})^4\rangle - 3$, where $\sigma$ and $\mu$ are the standard error and the mean value, respectively; and $\langle\rangle$ refers to the average over all data. Both $S$ and $K$ of a Gaussian distribution are equal to 0, and the positive values of $S$ and $K$ indicate the distribution curve has a more pronounced tail on the right side. From the inset of Fig.1, one can see that, as temperature decreases, $S$ and $K$ increase first, then reach maxima, and finally decrease with the decrease of temperature. Namely, at the high-energy side, the distribution deviates from the Gaussian form, and the relative deviation becomes larger toward low temperatures. For $Al_{43}$ and $Al_{46}$, the maximum values are achieved around 480 K and 520 K respectively, which are close to $T_g$.

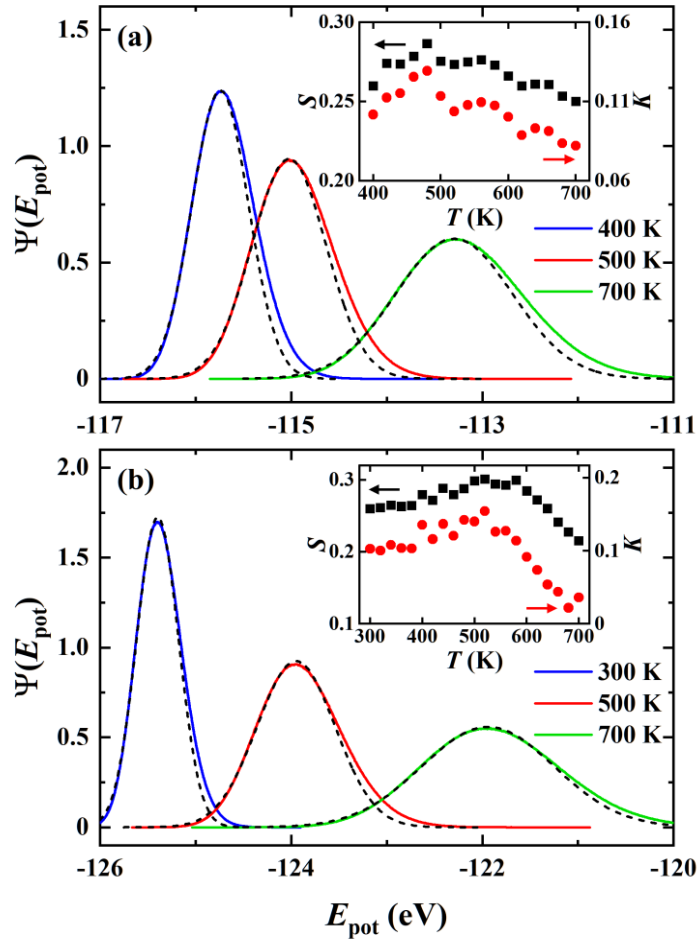

**Figure 1. The distribution of potential energy of $Al_{43}$ (a) and $Al_{46}$ (b). Dashed lines are Gaussian fits based on the data of low-energy regions. Inset: temperature dependence of skewness ($S$) and excess kurtosis ($K$) of the distribution $\Psi(E_{pot})$.**

The long-time distribution of potential energies should closely relate to the topologic structure of PELs, namely the distribution of local minima in PELs. However, abundant information about PELs is submerged in thermal fluctuation. As shown in the top panel of Fig. 2, the time evolution of $E_{\text{pot}}$ in $Al_{46}$ seems to show little difference at temperatures above, around and below $T_g$ (700 K, 500 K and 300 K, respectively) other than the amplitude of thermal fluctuations. By taking average of potential energy in a short period of time, such fluctuations can be partially smoothed, so the difference between high and low temperatures is gradually revealed. The evolution curves of $E_{\text{ave}}$ in the same period are presented in middle ($\Delta t = 1$ ps) and bottom panel ($\Delta t = 10$ ps) of Fig. 2. At higher temperature, *e.g.*, 700 K (green lines in Fig. 2), the system involves more possible relaxation processes. Thus, there exists a remarkable but uniform fluctuation in $E_{\text{ave}}$ even with longer average time ($\Delta t = 10$ ps, bottom panel of Fig. 2). However, at lower temperature, *e.g.*, 500 K, as $\Delta t$ increases over the time scale of thermal vibration (~1 ps), the time evolution of $E_{\text{ave}}$ clearly displays many plateaus with different heights and durations (red lines in Fig. 2). Such behavior can be observed at even lower temperature but is much rarer (*e.g.*, at 300 K, blue lines in Fig. 2, plateau appears around 1250 ps).

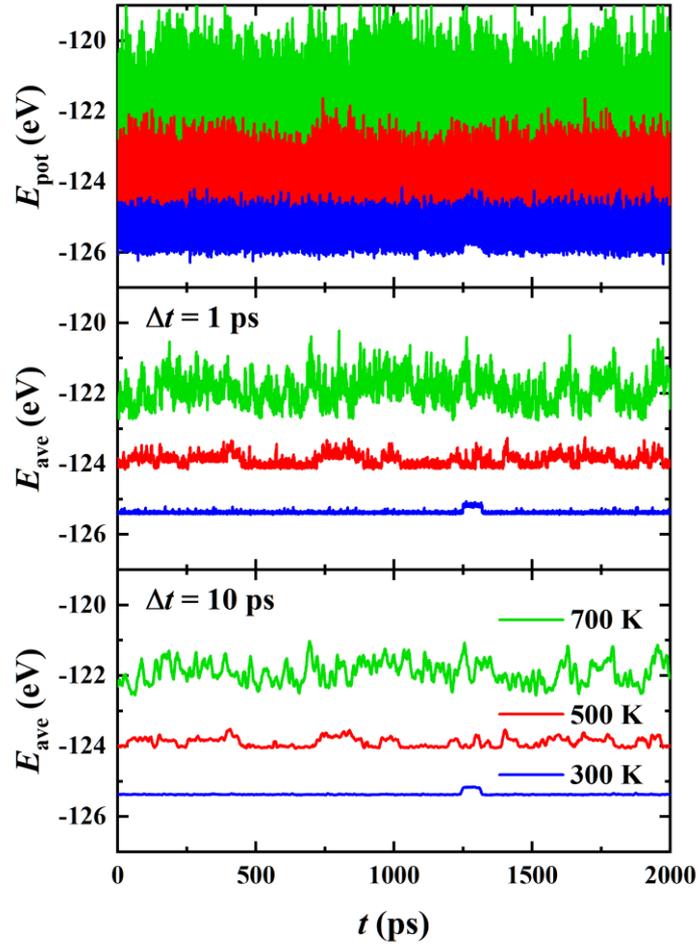

**Figure 2.** Time evolution of potential energy $E_{pot}$ of Al$_{46}$ (top panel) and its short-time average $E_{ave}$ (middle panel: $\Delta t = 1$ ps, bottom panel: $\Delta t = 10$ ps) in the same period. With the increase of $\Delta t$, at temperature of 500 K, $E_{ave}$ displays many plateaus with different heights and widths, which may reflect the inter-MB transition.

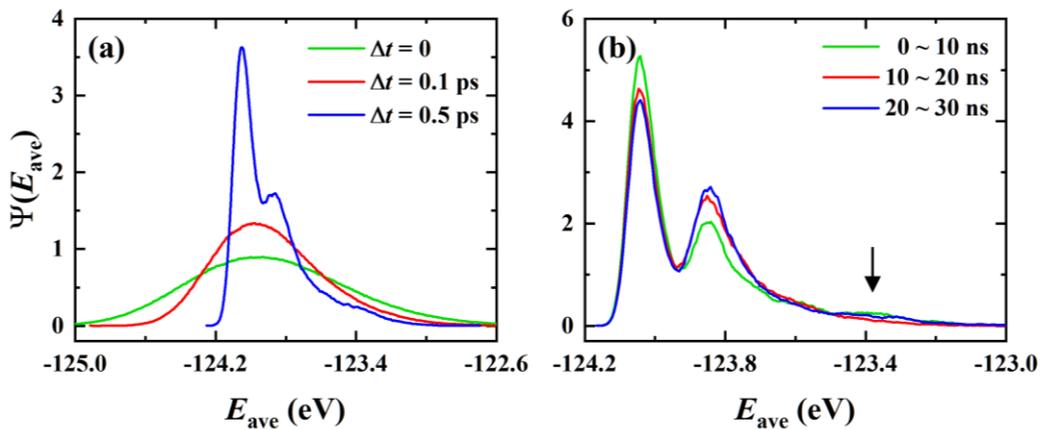

**Figure 3.** (a) The distribution of $E_{ave}$ of Al$_{46}$ at 500 K. With increasing $\Delta t$, the main peak splits into two, and a high-energy tail appears. (b) The distribution of $E_{ave}$ ($T$ = 500 K, $\Delta t = 1$ ps) in three adjacent intervals, the first 10 ns (green), the middle 10 ns (red), and the last 10 ns (blue). The arrow marks a less obvious sub-peak in the high-energy region at the first 10ns (green line).

Fig. 3(a) shows the distribution of $E_{ave}$ of Al$_{46}$ with different $\Delta t$ at 500 K. It can be seen that as $\Delta t$ increases the main peak in the distribution of $E_{ave}$ splits into two, together with the appearance of a high-energy tail. If $\Delta t$ increases further, the two Gaussian-like peaks become more obvious (Fig. 3(b), $\Delta t = 1$ ps). According to statistical physics, at equilibrium the energy should have a Gaussian distribution. The non-Gaussian distribution may be caused by the long relaxation time. When the relaxation time is comparable to the observation time, a true equilibrium distribution may never be achieved, thus a non-Gaussian distribution is expected. It may be the thermodynamic nature of nano glasses due to the very long relaxation time, which is similar to bulk glasses.[53,73,74]

The non-Gaussian distribution does vary with time, manifesting the fact that glasses have much longer relaxation time. Fig. 3(b) shows an example for Al$_{46}$ at 500 K with $\Delta t = 1$ ps. Firstly, both the heights and widths of the two main peaks change in different periods of time. Secondly, the tail in high-energy region in fact contains sub-peaks (marked by arrow in Fig. 3(b)). These sub-peaks are less distinguishable due to the smaller probability and wider distribution. Thirdly, the distribution of $E_{ave}$ is actually a combination of several Gaussian-like peaks. To be specific, a higher peak and several lower peaks on the right side are presented. The positions of these peaks correspond to the heights of the plateaus in Fig. 2, which are in fact the depth of the MBs (further discussed in the following context).

The paired-Gaussian and long-tailed distribution of potential energies only exists at supercooled liquids and glasses, which can be seen from Fig. 4. When temperature is much low (Fig. 4(a) for 200K) or much high temperature (Fig. 4(c) for 700K), the distribution of $E_{ave}$ exhibits the Gaussian form. And the increase of $\Delta t$ only changes the distribution width rather than the Gaussian form. The physical origin lies on the fact that, at much low temperature only the vibrations of atoms with typical timescale of picoseconds are excited, and at much high temperature the typical structural relaxation time is also in the order of picoseconds. Thus, comparing to the observation time, the relaxation time in both cases is much shorter, a true equilibrium distribution is achieved. However, in the intermediate temperature region, (Fig. 4(b)), the distribution of potential energies is clearly the non-Gaussian form in the whole range. By increasing $\Delta t$, the paired-Gaussian and long-tailed distribution of potential energies emerges.

The paired-Gaussian and long-tailed distribution of potential energies provides us a thermodynamic perspective to understand glass transition behavior at nano systems. A typical solid-liquid transition is accompanied by the generation of latent heat. It reflects the entropy difference between solid and liquid at the melting temperature, which is mainly due to the change in configuration entropy. For a conventional solid, the system is confined to a single potential minimum, accordingly the configuration entropy is zero. Thus, from liquid to solid there is a sudden change in the configuration entropy. Despite our current focus on nano glasses, if the paired-Gaussian and long-tailed distribution of potential energies also reflects the intrinsic characteristic of bulk glasses, we can explain why there is a residual entropy in bulk glasses.

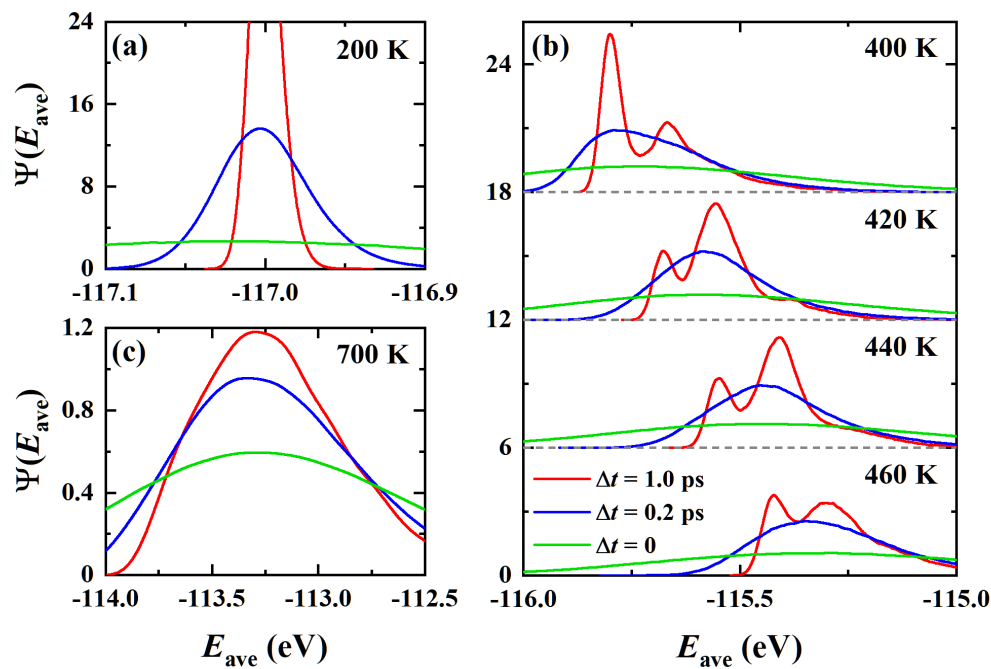

**Figure 4: Distribution of the short-time average of potential energy ($E_{ave}$) of Al$_{43}$ for much low temperature (a), intermediate temperatures (b), and much high**

temperature (c). For clarity, the curves of 440K, 420K and 400K in (b) has been offset vertically by 6, 12 and 18, respectively. For $T = 200\ K$ (a) and 700K (c), the increase of $\Delta t$ only decreases the distribution width. For intermediate temperatures (b), with the increase of $\Delta t$, the main peak starts to split into two peaks, and a long tail emerges at high energies.

## 2) Evidence of MBs

Below $T_g$, the variation in $E_{\text{ave}}$ clearly manifests the intra- or inter-MB transitions. For glasses, there are two types of relaxations, namely $\beta$- and $\alpha$-relaxation. With increasing $\Delta t$, information about $\beta$-relaxation, probably corresponding to transitions intra-MB, starts to blur first (middle panel of Fig. 2). However, with even larger $\Delta t$, wide plateaus remain, which should be related to $\alpha$-relaxations.

By comparing the evolution of $Q(t)$ and $E_{\text{ave}}$, we find that $E_{\text{ave}}$ is highly correlated with $Q(t)$. Fig. 5 shows the evaluation of $E_{\text{ave}}$ and $Q(t)$ in an arbitrarily selected time interval. Clearly the change in $E_{\text{ave}}$ and $Q(t)$ occurs synchronously. This connection between thermodynamic properties and structural transformations indicates that the remaining plateaus in $E_{\text{ave}}$ correspond to the average potential energy in MBs, and the change in $Q(t)$ manifests the associated structural transformations, *i.e.*, similar to the inter- or intra-MB transitions of bulk glassed.[11,13,75-77] The duration of the plateau corresponds to the residence time of the system in the local minima of PELs. When the duration is relatively long, it may correspond to an inter-MB jump. Detailed studies show that the structural transformation of inter-MB jump involves the cooperative motion of atoms, which has been observed in previous studies on both nano and bulk glasses.[12,13,68,78-83]

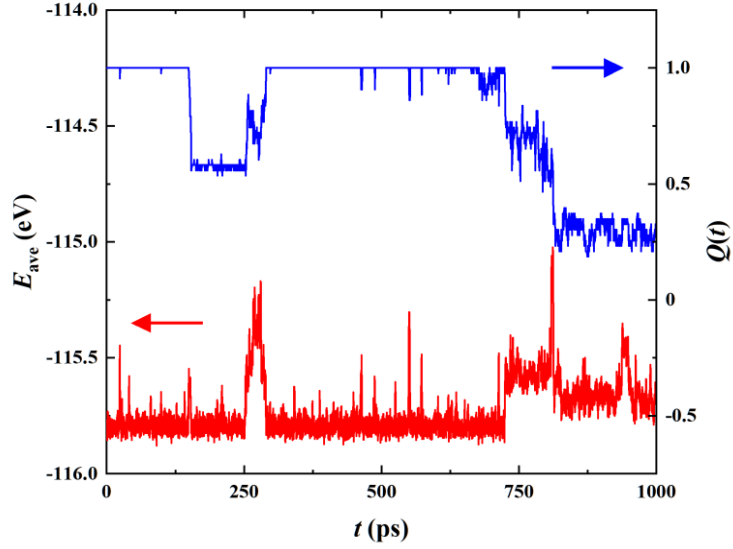

**Figure 5.** The short-time average of potential energy ($E_{ave}$) and the overlap function ($Q(t)$) of $Al_{43}$ ($T = 400K$, $\Delta t = 1ps$). $E_{ave}$ is highly correlated with $Q(t)$, indicating that the jump in $E_{ave}$ corresponds to structural transformations (jump in $Q(t)$).

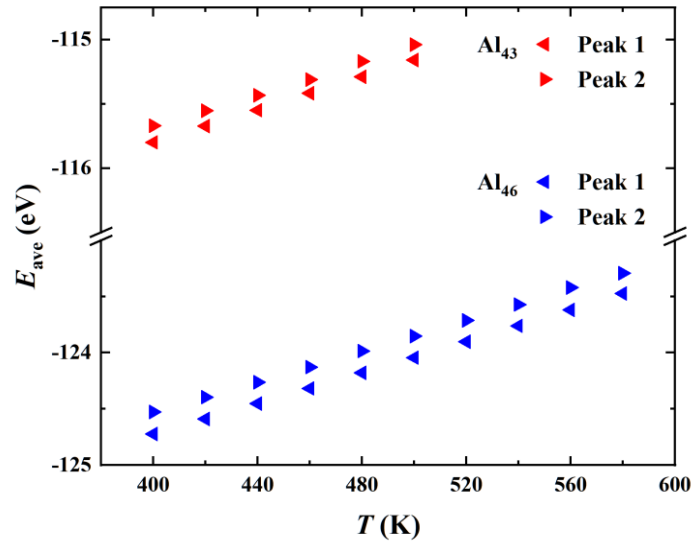

**Figure 6.** The central position of the two main peaks in the distribution of $E_{ave}$ for $Al_{43}$ and $Al_{46}$ at different temperatures. Left and right triangles represent the first and second peak, respectively. The distance between two main peaks is essential, almost independent with temperature.

Although the positions of the two main peaks in Fig. 3(b) and Fig. 4(b) vary with temperature and time, the difference between the two main peaks seems to be unchanged. To verify this issue, we have plotted the central position of the two main peaks in Fig. 6. It can be seen that the central position of the two peaks increases nearly linearly with temperature. The difference between the two main peaks stays almost unchanged, which is around 0.12 eV and 0.20 eV for $Al_{43}$ and $Al_{46}$, respectively. In Fig. 6, the data is not available at higher temperatures, since the bimodal structure is less distinguishable at high temperature.

The nearly constant difference between two main peaks may have its roots in the intrinsic nature of the PELs of nano glasses. There are a few similar two-level or two-state models, in which the transition between the two energy levels or states determines the thermal properties of the bulk glasses. The present results show that the two-level or two-state models[21,22,27,84] can indeed reflect the main characteristics of the nano glasses. However, what we would like to address is that, besides the two main peaks, there is a long tail in the high energy region. To our best knowledge, such a distribution has never been reported before, even in nano glasses. The long tail or sub-peaks at high energies is also important, and could make a non-negligible contribution to thermodynamic properties. One should take it into account when building thermodynamics models.

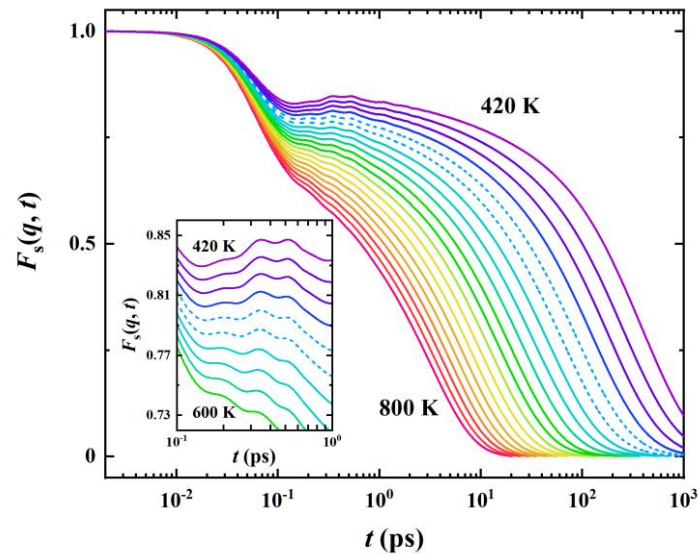

**Figure 7. Self-intermediate scattering function (SISF) at different temperatures of $Al_{43}$. The valley in the $\beta$ region emerges around $T_g$ (dashed blue curves) and**

**becomes deeper with decreasing temperature. From top to bottom, the temperature increment is 20 K.**

### 3) The *α*-relaxation

The paired-Gaussian distribution of $E_{ave}$, especially the high energy tail, is closely related to the long-time structural relaxation. SISF provides rich information on structural relaxations, which is shown in Fig. 7. From this figure, several interesting features can be seen. Above $T_g$, SISF has typical liquid-like (or supercooled-liquid) characteristics, which is consistent with previous results. However, below $T_g$, there exists an obvious valley between 0.1 and 1.0 ps, which has been found in network glasses but not in metallic glasses.[85] It is around $T_g$ that the valley emerges (for details see the insert of Fig. 7) and becomes deeper with the decrease of temperature. Combining with structure analysis, we find that the valley corresponds to the average residence time in a shallow minimum of the PELs, *i.e.*, *β*-relaxations. Once leaving this minimum, the system will transfer to other adjacent minima, which is indicated by the existence of a few following small valleys within 1 ps. For longer times, SISF decays rapidly, which corresponds to the inter-MB transition, *i.e.*, the *α*-relaxation.

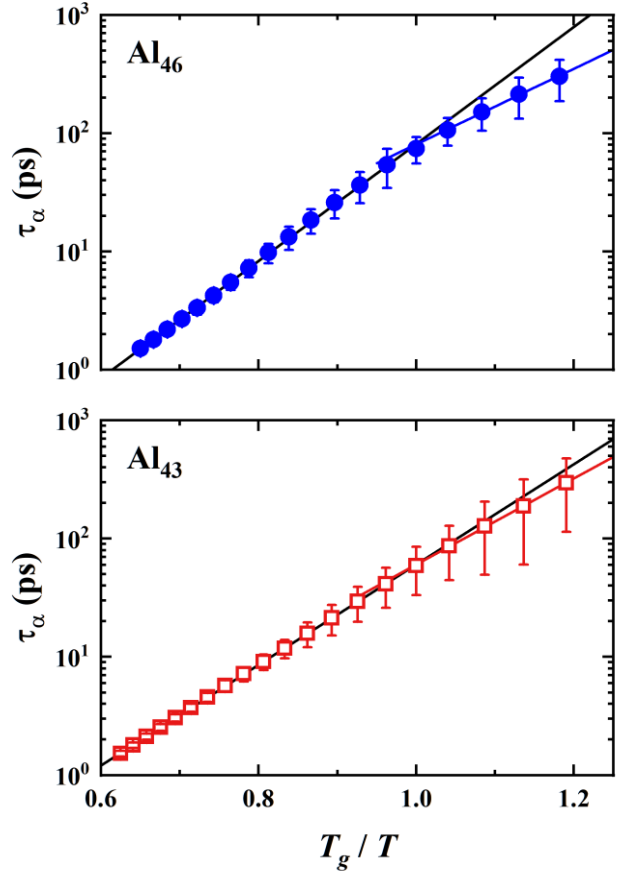

**Figure 8. Temperature dependence of $\tau_\alpha$. Below and above $T_g$, $\tau_\alpha$ curves can be well fitted by the Arrhenius equation separately, but with different activation energies. This behavior can be analogy with the well-known fragile-to-strong crossover in bulk glasses. The Arrhenius behavior indicates a strong glass. Interestingly the fluctuation grows to the same order of $\tau_\alpha$ below $T_g$.**

We can obtain the $\alpha$-relaxation time ($\tau_\alpha$), which is the time scale as SISF decays by $e^{-1}$. Fig. 8 shows $\tau_\alpha$ as a function of temperature. To our best knowledge, few MD simulations can obtain $\tau_\alpha$ with a pure dynamic meaning below $T_g$. From Fig. 8, one can see that, below and above $T_g$, $\tau_\alpha$ can be well fitted by two different exponential functions, namely the Arrhenius relationship. More importantly, the activation energies are lower below $T_g$, which is in agreement with previous calculation of diffusion constants[86]. Moreover, below $T_g$, the fluctuation of $\tau_\alpha$ grows much larger, which may indicate the temporal dynamic heterogeneity in the system.

The two Arrhenius plots can be analogy with the fragile-to-strong crossover reported in bulk glasses.[87-91] The fragile-to-strong crossover refers the structure relaxation changing from fragile (supper-Arrhenius) to strong (Arrhenius) behavior at the crossover temperature (usually slightly higher than $T_g$). The fragile-to-strong crossover found in current work is similar to that in bulk strong glass such as $SiO_2$ and $GeO_2$. For bulk strong glasses, the structural relaxation follows the Arrhenius relation for temperatures below and above the crossover temperature, but with different activation energies.[92] (Note: in some papers, the fragile-to-strong transition has the same meaning as the fragile-to-strong crossover[93], but in other paper it refers different phenomenon[60,94].) We believe some non-trivial characteristic structure of PELs can be found in varieties of glass formers in which similar fragility crossover exists. Interestingly, when the fragile-to-strong crossover occurs, both non-Gaussian parameters (the skewness ($S$) and excess kurtosis ($K$)) reach their maxima as shown in Fig. 1. The non-Gaussian distribution of potential energy may be the reflection of the breakdown of ergodicity, which is a consensus in the bulk glasses. This means that the fragile-to-strong crossover probably comes from the breakdown of ergodicity.

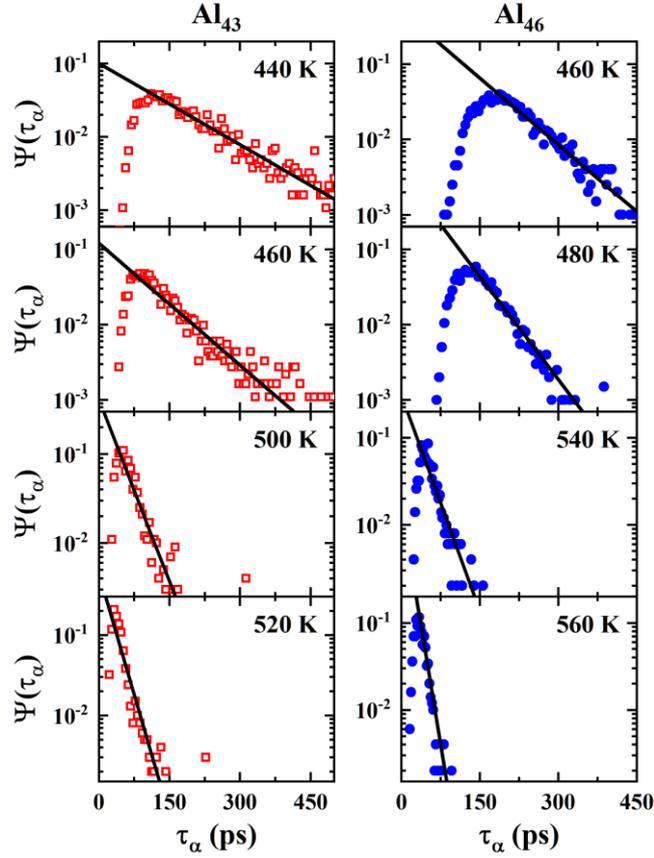

**Figure 9. The distribution of $\tau_\alpha$ of Al$_{43}$ (left panel) and Al$_{46}$ (right panel) at a few selected temperatures, where solid lines are the best fitting exponential decays. This exponential behavior may come from the fact that the inter-MB transition can really be regarded as a rare event.**

The distribution of $\tau_\alpha$ displays an exponential-like behavior for a few selected temperatures, which can be seen in Fig. 9. The exponential distribution here gives a reasonable explanation to our results shown in Fig. 8, in which the fluctuation of $\tau_\alpha$ below $T_g$ reaches the same order of magnitude as its mean value. It needs to be addressed that, when $\tau_\alpha$ is less than a certain value, the probability falls off quickly. It does not mean that the short-time relaxation process does not exist, or the exponential distribution breaks down. Probably, in short-time regions, $\alpha$- and $\beta$-relaxation, and even vibrations of atoms, are mixed together, which is really beyond the capacity of SISF for identifying the $\alpha$-relaxation. In a previous study of continuous-time random walks, the waiting time also shows a similar distribution.[95]

The above exponential distribution may be understood as follows. The inter-MB

transition occurs in a long-time scale below $T_g$, which can thus be regarded as a rare event. So, in the low temperature region, structural relaxation can be considered as a series of rare events occurring with a constant mean rate, whose behavior may be described by a Poisson-like process, and the probability of durations between events in a Poisson-like process follows an exponential distribution.[96]

It is of great interest to study the so-called stretched-exponential relaxation in these nano glasses, which has been recognized and widely discussed in bulk glasses.[97] For glasses, the correlation function could be well described by Kohlrausch-Williams-Watt expression:

$$\phi(t) = \phi_0 e^{-(t/\tau)^\beta}, \qquad (4)$$

where $\phi_0$ is a coefficient related to temperature, and $\beta$ is the so-called shape parameter. $\beta$, usual between 0 and 1, represents a universal feature of most glass-forming liquid, which indicates a stretched-exponential behavior comparing to a standard exponential relaxation. Recent works has also shown that in some bulk glasses, $\beta$ may be larger than 1 below $T_g$, indicating a compressed-exponential behavior [98,99].

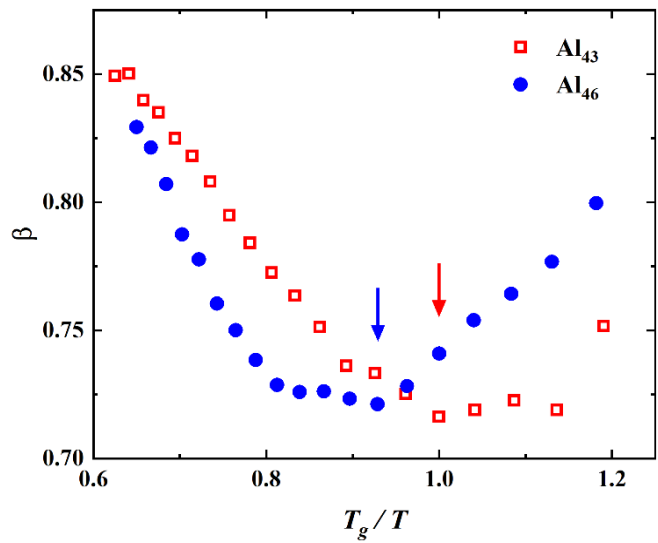

**Figure 10. Temperature dependence of $\beta$. In both Al$_{43}$ and Al$_{46}$ systems, a minimum occurs near the crossover temperature ($T_c$), which is marked with arrows. Above $T_c$, $\beta$ decreases with the decrease of temperatures; below $T_c$, $\beta$ increases with the decrease of temperatures.**

The temperature dependence of $\beta$ shows an unexpected feature. By fitting SISF curves using Eq (4), we have obtained $\beta$ at each temperature, which is shown in Figure 10. $\beta$ reaches a minimum near a certain temperature ($T_c$). Above $T_c$, $\beta$ decreases with the decrease of temperatures, indicating a more and more stretched-exponential behavior, which is in agreement with previous results[97]. In contrast, below $T_c$, $\beta$ increases with the decrease of temperatures. Here $T_c$ is in accordance with the crossover temperature for fragile-to-strong crossover. In consistent with the change of $\beta$, the distribution of relaxation time also changed significantly (Fig. 9). Compared with that above $T_c$, the distribution of relaxation time also changed significantly (Fig. 9). The distribution of relaxation time below $T_c$ becomes much wider than that above $T_c$. Since $T_c$ is around the fragile-to-strong crossover temperature, it implies that stretched-exponential behavior could also closely related to the fragile-to-strong crossover.

Various theories and models have been proposed to explain the novel stretched-exponential behavior. Although these models may root on quite different theoretical frameworks, they share one basic assumption: the relaxation time has a distribution ($\rho(\tau_\alpha)$) with a certain width.[97] Our data does show the existence of this kind of distributions. However, a strict calculation of $\rho(\tau_\alpha)$ requires to calculate $\tau_\alpha$ for each individual relaxation event, which seems to be impossible in MD simulations. In current work, to explore the distribution of relaxation time, the total simulation time is divided into a series of time intervals (~4 ns), and the average relaxation time is calculated in each time interval. Then the distribution of these averaged relaxation time is obtained, which is shown in Fig. 9. Since in each time interval, multiple relaxation events may occur, so the actual distribution ($\rho(\tau_\alpha)$) should in principle be wider than that shown in Fig.9. In fact, the SISF in each individual time interval could be stretched or compressed, but the averaged one always show the stretched-exponential behavior.

## Summary

Using MD simulations, we have systematically explored the PEL and structural relaxation in nano glasses. With an extremely long simulation (up to a few microseconds), we have observed the transition of systems among MBs. In contrast to the normal distribution, the distribution of potential energies becomes a non-Gaussian form, namely the paired-Gaussian and long-tailed distribution. The current study may provide a direct support for the widely adopted two-level or two-state models. However, the contribution from the long tail at higher energies should be properly considered. The relaxation time exhibits an exponential-like distribution, which results in the stretched-exponential behavior in $\alpha$-relaxation. We believe that the current results would also provide a useful reference for the study of bulk glasses.

**Acknowledgements**

Project supported by the National Natural Science Foundation of China (Grant No. 11874148). The computations were supported by ECNU Public Platform for Innovation.